# A simulation of the plasmonic absorption of the silica glasses with copper selenide nanoparticles

*V. S. Gurin*

*Research Institute for Physical Chemical Problems, Belarusian State University, Leningradskaya, 14, 220006 Minsk, Belarus*

*e-mail: gurin@bsu.by*

**Abstract**

The plasmonic optical absorption of the silica glasses containing copper selenide nanoparticles is simulated on the basis of Drude theory. The plasmonic resonance absorption is studied in dependence on plasmonic frequency, damping factor and the matrix dielectric function. The charge carrier concentration in the nanoparticles is evaluated through the plasmonic frequency for the spectra of closest correspondence to experimental. The plasmon resonance position and the width of maxima are varied throughout the visible and near-IR ranges for the above parameters of experimental glasses with $Cu_{2-x}Se$ nanoparticles.

## 1. Introduction

Plasmonics is well known was not created recently, however last decades, approximately from the beginning of XXI century this field of science acquired new resonance in optics of solid matter [1,2]. In spite of long history and intensive present studies a lot of novel physical effects continue to be discovered. These are: alternative plasmonic (i.e non-metallic) materials, epsilon-near-zero (ENZ) composites, a strong plasmon-matter coupling combinations, plasmon-activated chemistry and photocatalysis, etc. [1-3]. All that not only opens the deeper understanding plasmon physics but also lead to new advanced applications in different fields of technology and device construction. The present work concerns the area of nanoscale semiconductors those properties are provided by increased concentration of charge carriers approaching them to the plasmonic, metal-like materials. These are chalcogenides of copper, silver, ternary chalcogenides of the $A^I B^{III} C^{VI}$ composition. Also, many oxides: ZnO, $SnO_2$, $WO_3$ heavily doped by various elements, TiN, AlN, etc. Nanoscale semiconductors of these classes can form the structures of different dimensionality, 0D, quantum dots, wires and rods 1D, thin films and platelets, 2D, etc. The structures can be assembled into more complicated spatial ordered building blocks and these features can provide contributions into plasmonic properties and more physical effects.

The works acknowledged as pioneering on plasmonics for chalcogenide semiconductors [4-6] have reported the copper sulphides and selenides. The fabrication technique was taken as the development of the actively treated methods for II-VI and IV-VI semiconductors (CdSe, ZnS, PbS, etc). It is noticeable that both $Cu_xS$ and $Cu_xSe$ nanostructures (particles, films, matrix-embedded species) have been fabricated earlier, however, the optical features were not interpreted within the framework of plasmonic concepts. The alternatives to explain the NIR absorption band appearing for the copper chalcogenide nanoparticles in some cases were suggested as intraband transitions and/or surface states due to partial oxidation [7,8]. The size variation for particles and thin films thickness appear to be as additional factor to result in complicacy of optical features of these materials [9,10].

In the present work, we consider the plasmonic optical materials those are the sol-gel fabricated silica glasses with incorporated copper selenide nanoparticles [11,12] and perform the simulation of their optical absorption that appears due to the plasmonic resonance in the nanoparticles. These glasses are an example of the 'particles-matrix' nanocomposite system in which the matrix is an amorphous $SiO_2$ (prepared through the sol-gel technique) and the nanoparticles are semiconducting copper selenide particles stabilized in the silica matrix at considerable low concentration. The particle size range for them is 5-100 nm and can be controlled by the preparation protocol and the copper precursor concentration. The chemical composition of the particles may be described as $Cu_{2-x}Se$, $0<x<0.3$ and their crystalline structure is close to one of stable low-temperature copper selenide phase (berzelianite) [13,14]. Optics of the glasses with copper selinide is of special interest because explicit appearance of both size effects of quantum-confined semiconductors and the plasmon resonance for nanoparticles. In contrast to the colloidal chalcogenide nanoparticles those much more intensively treated, the glasses are more appropriate as non-linear optical media for laser technology due to higher mechanical and optical strength. Meanwhile, for these materials no simulation of plasmonic properties was done to date. Of course, in general concepts, the nature of the plasmonic light absorption is understandable, it is similar to the copper chalcogenides synthesized in colloids. In both cases, the principal factor providing the plasmonic features is an elevated concentration of charge carriers. Therefore, the purpose of this work is the evaluation of it on the basis of the Drude theory.

## 2. The simulation method

The absorption spectra of the plasmonic nanoparticles were calculated using the familiar expression for the absorption coefficient derived from the Mie theory for particles with small sizes $R \ll \lambda$ [15]

$$\alpha = \left(\frac{18\pi C \varepsilon_m^{3/2}}{\lambda}\right) \frac{\varepsilon_2}{(\varepsilon_1 + 2\varepsilon_m)^2 + \varepsilon_2^2} \qquad (1)$$

or, by transforming wavelength into frequency, as $\omega = 2\pi c/\lambda$, where $c$ is the light velocity:

$$\alpha = \left(\frac{9C\omega \varepsilon_m^{3/2}}{c}\right) \frac{\varepsilon_2}{(\varepsilon_1 + 2\varepsilon_m)^2 + \varepsilon_2^2} \qquad (2)$$

In this formula $\varepsilon_1$ and $\varepsilon_2$ are real and imaginary parts of dielectric function for the nanoparticles, $\varepsilon_m$ stands for the dielectric function of medium in which the particles are located. $\omega_p$ is the plasmon frequency that is determined within the framework of this model through free carriers concentration N as

$$\omega_p^2 = Ne^2/\varepsilon_0 m_{ef}, \qquad (3)$$

where $e$ is the elementary charge, $\varepsilon_0$ is the fundamental dielectric constant and $m_{ef}$ means the efficient mass of charge carriers. In the formula (1) C is a volume particle concentration in the medium. Note, the approach used assumes the low particle concentration and the interaction with light proceeds with separate particles. This is quite good approximation for the materials under study and further development is expected for similar glasses with higher particle concentration. To calculate the absorbance plots in arbitrary units in correspondence with conventional measurements with spectral devices we omit an explicit inclusion of the frequency-independent particle concentration.

Within the classical Drude theory for plasmons in nanoparticles $\varepsilon_1$ and $\varepsilon_2$ are expressed through the following formulas in dependence on frequency including also the damping factor $\gamma$:

$$\varepsilon_1 = \varepsilon_\infty - \frac{\omega_p^2}{\omega^2 + \gamma^2}; \quad \varepsilon_2 = \frac{\gamma}{\omega} \frac{\omega_p^2}{\omega^2 + \gamma^2} \qquad (4)$$

In the range of optical frequencies (UV-Vis-NIR, $\lambda < 2$ μm) the damping factor is small and these expressions become simpler:

$$\varepsilon_1 = \varepsilon_\infty - \frac{\omega_p^2}{\omega^2}; \quad \varepsilon_2 = \frac{\gamma \omega_p^2}{\omega^3} \quad (5)$$

The value $\varepsilon_\infty$ in these formulas is the "non-plasmonic" part of dielectric function associated basically with interband absorption and the interaction of incident light with crystalline lattice. Within the framework of this paper we avoid more detailed discussion of these contributions. They accounted through the bulk dielectric function of copper selenide, $\varepsilon_\infty$. Its value is 11.6 [16] (naturally, there are some variations in dependence of exact $Cu_xSe$ stoichiometry, surface state of nanoparticles, preparation technique, etc). Ref. 16 takes the summarized data, but the exact value itself does not influence the principal outcomes of the behavior of plasmonic response of nanoparticles.

The size effects for the plasmonic absorption of nanoparticles simulated through the Drude theory may be included (one of simplest way) as the influence of finite particle dimension upon the damping factor [15] $\gamma = \gamma_{bulk} + V_f/L$, where $\gamma_{bulk}$ is the damping of the bulk semiconductor, $V_f$ stands for the velocity of electrons at the Fermi surface, and L accounts the geometry of particles, for spheres with radius R it can be given as $L = 4R/3$. Therefore, the calculation of absorption spectra can be performed through the above formulas taking the value of plasmonic frequency determined by charge carrier concentration, the damping factor and the properties of medium in which the particles are localized through $\varepsilon_m$.

## 3. Results and Discussion

The plasmon frequency, $\omega_p$, is the principal parameter providing the feature of plasmonic resonance for nanoparticles. Its effect clearly depicted in Fig. 1. Little change of $\omega_p$ results in the rather significant effect: the less value of the plasmon frequency shifts the spectrum to the right side accompanied by fall of intensity and strong broadening. Therefore, this key factor in the range of $(0.5 \div 1.2)*10^{16}$ с$^{-1}$ can cover the interval of maxima for experimental spectra of the glasses with plasmonic copper selenide nanoparticles.

The value of $\omega_p$ can be used to estimate the carrier concentration that is responsible for the plasmon resonance generation in these particles according to the formula (3). For the spectrum related to $\omega_p=10^{16}$ c$^{-1}$ one can derive $\cdot 5.9\cdot 10^{21}$ cm$^{-3}$. We use the value of hole effective mass 0.23 $m_e$ [14] (however, it should be noted that there are variations in the reliable data for this type of materials) and this estimated carrier concentration fits quite good to the data for copper selenide nanoparticles produced in colloidal state, $10^{21}$ -$10^{22}$ cm$^{-3}$ [17,18].

Fig. 2 displays a series of calculated absorption spectra for the copper selenide nanoparticles for variable dielectric function of the surrounding medium, $\varepsilon_m$. The medium is assumed close to silica, and this value was tested in the range of 1.5-2.5. These spectra demonstrate the red shift and growth of intensity for the higher $\varepsilon_m$. The effect is shown to be rather significant that opens possibility to design plasmonic optical sensors on the basis of this phenomenon [19].

The next factor affecting the plasmon resonance of nanoparticles under study is the damping, $\gamma$. As noted above, it may be associated directly with size (and geometry) of nanoparticles, however, another contributions those are out the scope of the present publication could be also expected: defects, surface states, temperature, etc. [20]. A series of spectra in Fig. 3 indicate clearly that the higher damping provides the plasmon resonance of lower intensity and the maxima shifted to the shorter wavelengths. Evidently, the more damping factor can make the resonance difficult to observe. That is the reason its absence for complex experimental situations when the model considered here can be violated.

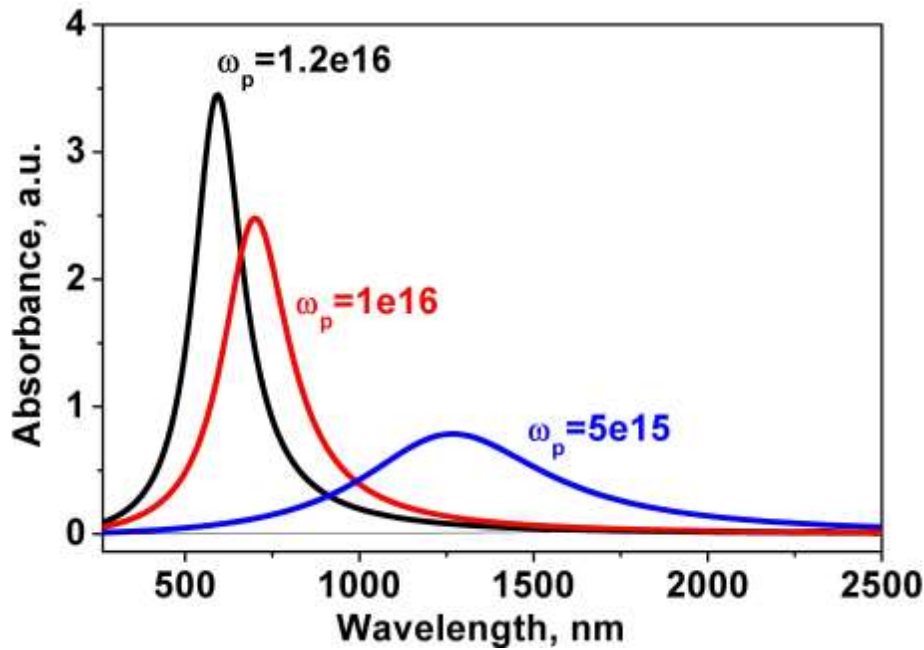

Fig. 1. The simulated absorption spectra of the glasses with copper selenide nanoparticles for different values of plasmonic frequency $\omega_p$

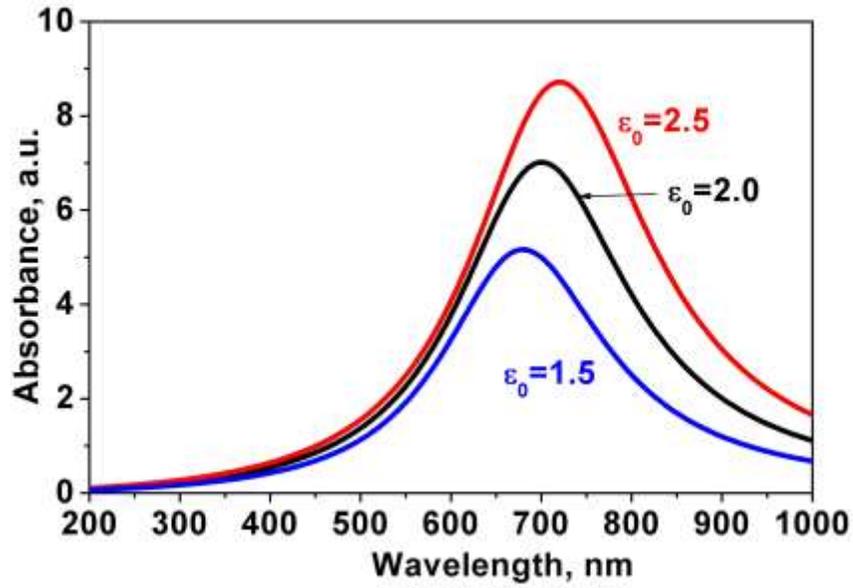

Fig. 2. The simulated absorption spectra of the glasses with copper selenide nanoparticles for different values of the matrix dielectric function, $\varepsilon_m$.

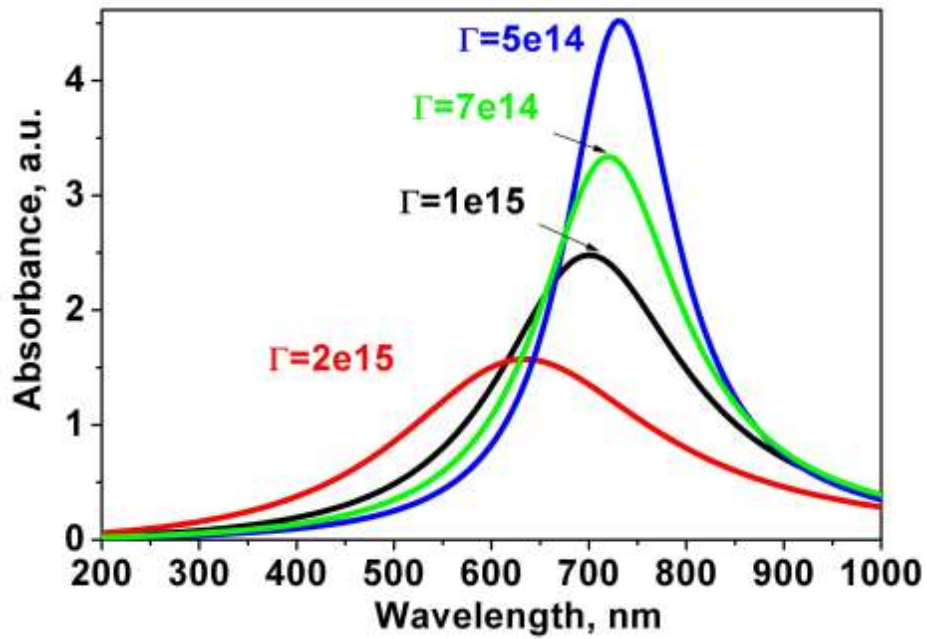

Fig. 3. The simulated absorption spectra of the glasses with copper selenide nanoparticles for different values of the damping factor, $\gamma$.

## Conclusions

The plasmon resonance phenomenon is simulated for the semiconductor nanoparticles possessing elevated charge carrier concentration on the basis of Drude theory for the electron gas in nanoparticles. The theory is implemented for the sol-gel derived glasses with copper selenide nanoparticles. The plasmon resonance maxima are shown to be varied in the visible and near-IR interval (400-1200 nm) under the change of plasmonic frequency, damping factor and the matrix dielectric function within the ranges corresponding to the sol-gel derived glasses with $Cu_{2-x}Se$ particles with x close to 2. From the values of plasmonic frequency the charge carrier (holes) concentration is estimated about $5.9*10^{21}$ $cm^{-1}$. Thus, the simple Drude model can be used for interpretation of the principal features of the silica glasses with plasmonic copper selenide nanoparticles.

## Acknowledgement

The work was performed under support of the State Program of Scientific Investigations of Belarus "Nanostructure" (2021-2025).